# Framework for A Personalized Intelligent Assistant to Elderly People for Activities of Daily Living

**Nirmalya Thakur**  *thakurna@mail.uc.edu*
*Department of Electrical Engineering and Computer Science*
*University of Cincinnati*
*Cincinnati, OH 45221-0030, USA*

**Chia Y. Han**  *han@ucmail.uc.edu*
*Department of Electrical Engineering and Computer Science*
*University of Cincinnati*
*Cincinnati, OH 45221-0030, USA*

**Abstract**

The increasing population of elderly people is associated with the need to meet their increasing requirements and to provide solutions that can improve their quality of life in a smart home. In addition to fear and anxiety towards interfacing with systems; cognitive disabilities, weakened memory, disorganized behavior and even physical limitations are some of the problems that elderly people tend to face with increasing age. The essence of providing technology-based solutions to address these needs of elderly people and to create smart and assisted living spaces for the elderly; lies in developing systems that can adapt by addressing their diversity and can augment their performances in the context of their day to day goals. Therefore, this work proposes a framework for development of a Personalized Intelligent Assistant to help elderly people perform Activities of Daily Living (ADLs) in a smart and connected Internet of Things (IoT) based environment. This Personalized Intelligent Assistant can analyze different tasks performed by the user and recommend activities by considering their daily routine, current affective state and the underlining user experience. To uphold the efficacy of this proposed framework, it has been tested on a couple of datasets for modelling an "average user" and a "specific user" respectively. The results presented show that the model achieves a performance accuracy of 73.12% when modelling a "specific user", which is considerably higher than its performance while modelling an "average user", this upholds the relevance for development and implementation of this proposed framework.

## 1. INTRODUCTION
The current century has seen a rapid increase in the population of elderly people [1] and it is predicted that their population is to even exceed the number of children within a few decades [2,3]. This rapid increase in the number of elderly people, mostly characterized by the increase in population of young elderly (aged between 65 to 85 years) as shown in Figure 1, is primarily due to improved conditions of health, assisted living facilities and declining fertility across the world [2]. Increasing age, which is characterized with the increasing needs for caregiver and healthcare solutions is starting to become a burden on the world's economy [4]. The number of elderly people across the world with dementia has doubled in recent times [5] and their number is predicted to again double by the year 2030, leading to approximately 76 million people with dementia worldwide. In 2010 alone, approximately $604 billion costs were incurred to the healthcare industry in looking after people with dementia and this number is increasing at an alarming rate. [5]

Nirmalya Thakur & Chia Y. Han

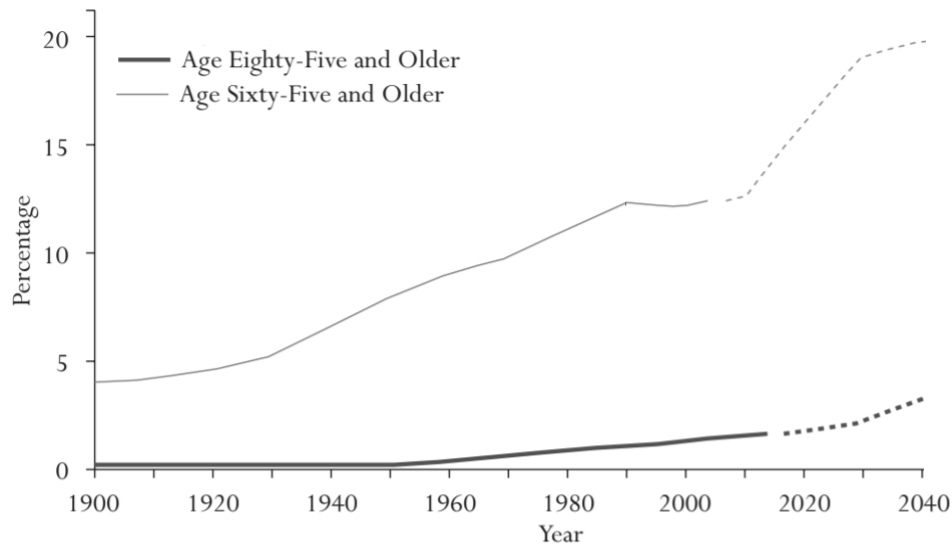

**FIGURE 1:** The current and predicted increase of elderly people: aged 65-85 and aged 85+ compared. (based on data about the population of elderly people from World Health Organization [9]).

For technology to sustain this ever-increasing population of elderly people and address their requirements, impactful and forward-looking urban development policies by governments and sustainable infrastructures implemented with technologies, for example smart homes, are necessary to support and enhance the quality of life experienced by them.

The application of Affect Aware Systems in an Internet of Things (IoT) based smart and connected environment holds the potential to serve as a long-term feasible solution to address this challenge. In the context of a smart home, Affect Aware Systems may be summarized as intelligent systems that can analyze the affective components of user interactions with an aim to assist the user to perform daily activities and improve their user experience in the context of their day to day goals. To enhance the quality of life experienced by elderly people and reduce their anxiety towards adapting to new technologies, it is essential for Affect Aware Systems to not only analyze user behavior [6] but also to recommend activities based on affective states [7] and the underlining user experience, to create a reliable, assistive, trustworthy and context-aware environment for assisted and ambient living.

The essence of providing such technology-based solutions lies in the effectiveness of technology to address the diversity in elderly population which can be broadly characterized by their varying age group, gender and differences in their background [8]. This diversity in the elderly population leads to varying experiences resulting in different habits and diverse nature of user interactions. Elderly people can broadly be subdivided into two sub-groups based on their age – (1) Young elderly – aged 65 to 85 and (2) Old elderly - aged 85 and above. These specific sub-groups are also diverse in terms of their gender and recent research [9, 10] has shown that this also leads to different characteristics of user interactions in elderly people.

Recent researches [11-20] in the field of affective computing and human-computer interaction for improving the quality of life of elderly people by providing cost effective solutions to address their needs have mostly focused on modelling an "average user", however there is quite often a gap between the "average user" for which a specific technology is proposed and the "actual user" that exists - thereby leading to ineffectiveness of the assistive technology to aid the user. Therefore, this paper proposes the framework for development of a Personalized Intelligent Assistant that can adapt according to the user interactions performed by any "specific user" and recommend activities based on daily routine, current affective state and the underlining user experience, to improve the quality of life experienced by elderly people in the context of their day to day activities in a smart home.

Nirmalya Thakur & Chia Y. Han

This paper is organized as follows: Section II provides an overview of related works in this field. Section III provides details about the proposed framework which is followed by Section IV which discusses the results and findings. The conclusion and scope for future work is presented in Section V which is followed by references.

## 2. RELATED WORK

This section reviews the recent researches in this field which have focused on activity recognition and activity recommendation for creation of assistive living spaces in the context of smart home environments. Activity recognition models can broadly be classified into two categories – knowledge-based models and data driven models. Knowledge based approaches mostly make use of ontologies to infer about activities. Azkune et al. [11] proposed a multilayered framework to analyze human activities in the context of a smart home. The architecture consisted of multiple layers which were associated with the tasks of acquisition of the data collected from wireless sensors, understanding the semantics, and providing descriptive knowledge of the recognized action or task.

In the work done by Riboni et al. [12], ontologies were not used to infer the specific activity being performed but they were used to validate the result inferred by the activity recognition model. The model developed a knowledge base of tasks associated with different context parameters in a smart home environment and used a combination of probabilistic reasoning and statistical modeling to analyze different activities. Nevatia et al. [13] developed a formal language, based on image recognition, for analyzing activities based on video recordings and real time streaming video data.

Data driven approaches for activity recognition have typically involved implementation of machine learning and data analysis methods. Kasteren et al. [14] developed an activity analysis model that involved representing different activities using each state of a Hidden Markov Model (HMM). The model had features to understand the raw data coming from motion sensors and analyze the rate of change of the instantaneous readings of this raw data, to infer about the given activity being performed by the user.

Cheng et al. [15] developed a hierarchical model with three layers to perform activity recognition based on video content analysis using multiple kernel learning methods. This approach was able to analyze both individual and group activity based on motion information by taking into consideration action trajectories and the information about effect of these actions on the context parameters. Skocir et al. [16] developed a system consisting of infrared sensors based on Artificial Neural Networks (ANNs) for activity recognition. The system was able to infer enter and exit events in different rooms in a smart home based on sensor information.

A context aware task recommender system was developed by Doryab et al. [17] to augment practitioners' performances in a specific hospital environment. The system detected the course of actions performed by the user at a given point of time and recommended tasks that were associated to these courses of actions from a knowledge base. A recent work by Thakur et. al [18] proposed a framework that analyzes multiple activity instances performed by different users, to provide a general definition for the given activity in the specific environment. This work also proposed an activity recommendation system that could identify distractions and recommend tasks to the user as per this general definition of an activity in the given environment. An unsupervised recommender system was proposed by Rasch [19] that analyzed the habits of the users and communicated with other systems to create habitable user experiences. A recommender system in the domain of healthcare was developed by Vavilov et al. [20] to recommend different tasks to patients to aid their recovery. A recommender system based on individual profiling method was proposed by Mark C et al. [21]. It develops individual profiles in its memory for all the users to be able to recommend tasks to individuals in a better way. Gong et al. [22] analyzed social media information of users to gather more information about the non-verbal aspects of user interactions to recommend better tasks. A recommender system based out of content filtering and neighborhood based collaborative filtering was proposed by Lai et al. [23].

Nirmalya Thakur & Chia Y. Han

Majority of the works [11-16] done in this field have focused on activity recognition by various methodologies. The few works [17-23] that have focused on developing activity or task recommender systems, have mostly taken into consideration user interactions of multiple users to define an "average user" and recommend tasks or actions based on the same. However, in a realistic scenario, the traits and characteristics of this "average user" might be significantly different from a "specific user" in the given context, owing to the user diversity. This could lead to failure of such systems to effectively assist users in the given environment and lead to barriers in the context of fostering human-technology partnerships. Therefore, the need to provide a technology-based solution that can adapt to user interaction patterns of any "specific user" and create an assistive environment for elderly people to improve their quality of life and augment their performances in the context of their day to day goals, is highly necessary. This serves as the main motivation for this work.

A couple of related works that have been used to propose this framework are (1) A Complex Activity Recognition Algorithm (CARALGO) for analyzing human behavior [24] and (2) A Complex Activity Based Emotion Recognition Algorithm (CABERA) for Affect Aware Systems [25] to analyze the emotional response associated to different complex activities.

According to CARALGO [24]. any complex activity (WCAtk) can be broken down into small actions or tasks – these are called atomic activities (At) and the context parameters that affect these atomic activities are called context attributes (Ct). Each of these atomic activities and context variables are associated with specific weights based on probabilistic reasoning. Each complex activity has a set of specific atomic activities that are essential for performing the activity – these are called core atomic activities (γAt) and the context parameters affecting them are called core context attributes (ρCt). Based on the weights of atomic activities and context attributes associated to the complex activity, every complex activity is associated with a threshold function (WTCAtk) that helps to determine the occurrence of that activity. The total weight for any given occurrence of this complex activity should be equal to or greater than the value of its threshold function for the complex activity to have been successfully performed. In the event when the weight is less than the value of the threshold function, it helps to infer that the activity was not completed successfully by the user which could be due to several factors. CARALGO also helps to identify the start atomic activities (AtS), start context attributes (CtS), end atomic activities (AtE) and end context attributes (CtE) related to a complex activity.

CABERA [25] helps to analyze the emotional response of different complex activities in the context of a smart home based on the probabilistic analysis of complex activities using atomic activities and their associated context attributes. It starts with analyzing the condition for occurrence of the complex activity by checking for the threshold condition. Thereafter the atomic level analysis of the complex activity is performed to identify the most important atomic activity and the associated most important context attribute. Then probabilistic reasoning principles are applied to analyze the nature of occurrence of this complex activity at different time instants. This is done by studying the nature of the most important atomic activity and the associated most important context attribute over these time instants when the activity occurred. This helps in drawing an inference about the emotion (positive or negative) associated to the complex activity at the given time instant.

## 3. PROPOSED WORK
Implementation of this proposed framework for development of a Personalized Intelligent Assistant for recommending Activities of Daily Living (ADLs) to elderly people in a smart home comprises of the following steps:
1. Develop a database of user interactions in the context of day to day activities and ADLs in a smart and connected IoT-based environment.
2. Determine interesting characteristics of complex activities in terms of their context parameters. This involves analyzing the appliance usage patterns in the context of a given complex activity.

Nirmalya Thakur & Chia Y. Han

3. Analyze multiple activity instances based on daily routine and investigate the time sequence of multiple activities to identify typical macro activities.
4. Use CARALGO to analyze the multimodal aspects of user interactions associated to these macro activities by extracting the atomic activities, context attributes, core atomic activities, core context attributes, start atomic activities, start context attributes, end atomic activities and end context attributes.
5. Use CABERA to deduce relevant patterns of these activity occurrences that provide indication about the affective state of the users performing these activities.
6. Use a supervised learning approach to relate the emotional response of users to user experiences associated to these activities.
7. Develop a supervised learning approach to implement a recommender system that can recommend complex activities based on these patterns, affective states and the associated user experience.

This proposed framework was implemented to first model an "average user" and then model a "specific user" from two different datasets to observe the differences in its performance and working in the two different scenarios, which is presented next. This implementation was done in RapidMiner [28]. RapidMiner is a data science software platform which provides an integrated development environment for implementation of data analysis, machine learning, deep learning and natural language processing algorithms.

RapidMiner is developed on an open core model which provides a GUI to enable users to execute workflows which are defined as "processes" in RapidMiner repository. These "processes" can be simulated by connecting multiple "operators" in a logical sense and as per the requirement of the given "process". Each "operator" is RapidMiner is associated with the basic definition of a specific task or function which can be modified each time by the user as per the requirement. There are currently two versions of RapidMiner available – the free version and the commercial version. For implementation of this framework as discussed in the subsequent sections, the free version of RapidMiner [28] was used.

**3.1 Implementation of the Framework to model an "average user"**
This involved analyzing activities from a subset of the UK Domestic Appliance Level Electricity (DALE) dataset [27]. The UK DALE dataset consists of details about appliance usage patterns related to different complex activities, measured with a time resolution of 6 seconds, in five different smart homes in Sothern England, recorded over a period of three years from 2012 to 2015.

These appliance usage patterns have been analyzed to obtain information about the different complex activities performed in different smart homes. Figure 2 shows different instances of occurrences of these complex activities from this dataset.

Two specific activity states, according to CARALGO are analyzed in this approach. For each time instant, '1' represents the fact that the activity was performed and '0' represents the fact that the activity was not performed. The analysis involved studying the occurrences of multiple instances of the same activity over specific time intervals to study their associated patterns.

Thereafter the atomic activities and context attributes associated to all these complex activities were analyzed. The different instances of occurrences of the complex activities – Using Washing Machine and Cooking in Kitchen are shown in Figures 3-4. The CARALGO analysis of all the complex activities – Watching TV, Using Laptop, Using Subwoofer, Using Washing Machine, Cooking in Kitchen, Using Microwave and Using Toaster are shown in Tables 1-7.

Nirmalya Thakur & Chia Y. Han

This analysis by CARALGO involved identifying the actions or tasks performed by the user while doing the given complex activity along with studying the context parameters on which these actions or tasks were performed. The process then involved associating weights to these

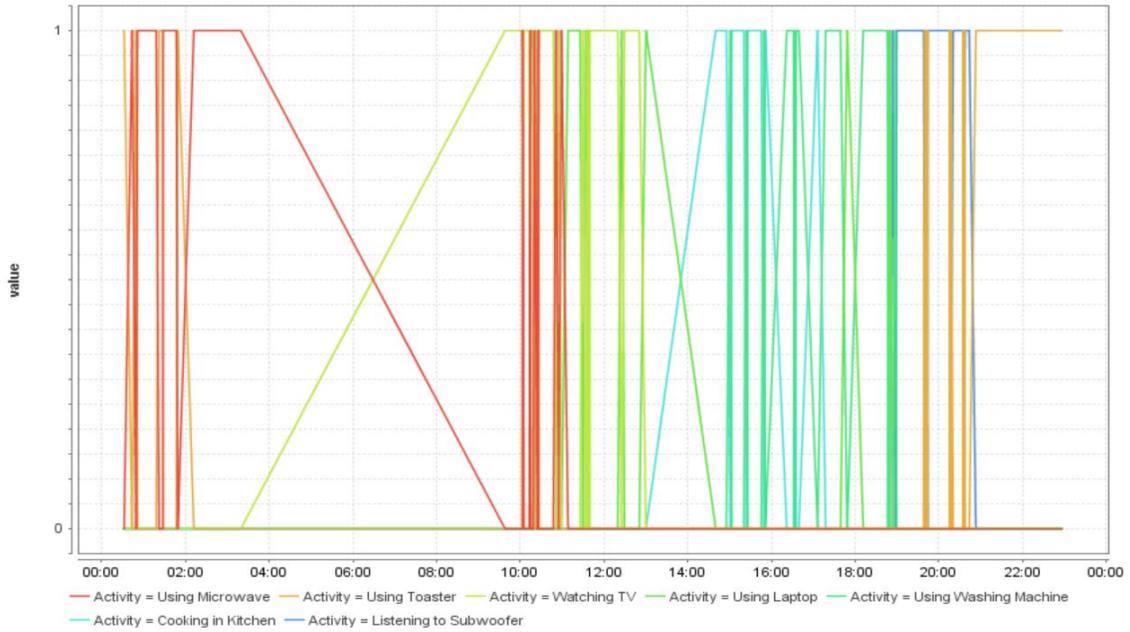

**FIGURE 2:** Different instances of activity occurrences from the UK DALE Dataset. This includes the complex activities of Watching TV, Using Laptop, Listening to Subwoofer, Using Washing Machine, Using Microwave, Cooking in Kitchen and Using Toaster.

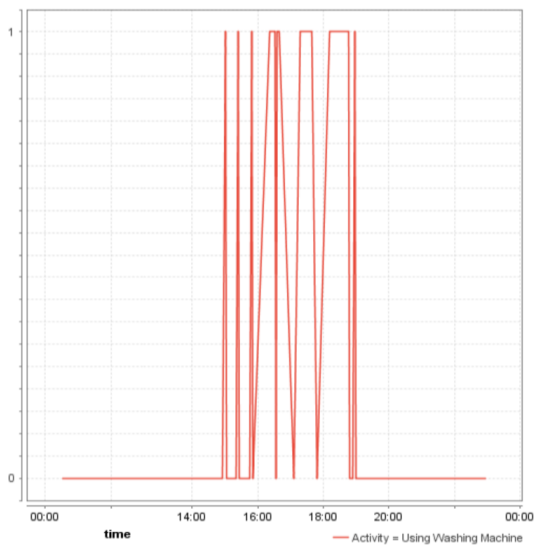 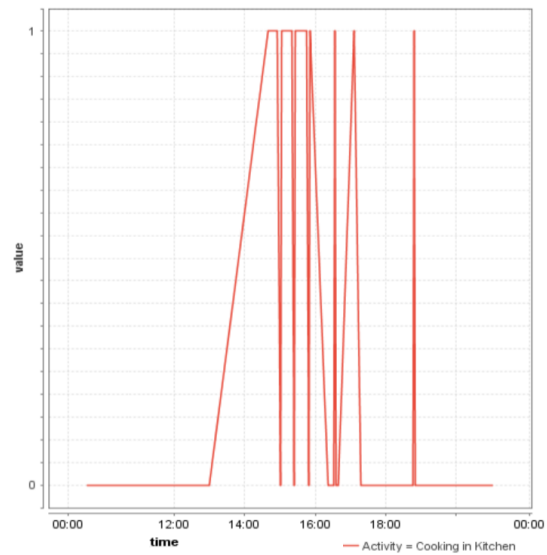

**FIGURE 3:** Multiple instances of the complex activity of Using Washing Machine.

**FIGURE 4:** Multiple instances of the complex activity of Cooking in Kitchen.

Nirmalya Thakur & Chia Y. Hanrespective atomic activities and context attributes based on probabilistic reasoning. Based on the weights of these atomic activities and context attributes the core atomic activities, core context attributes, start atomic activities, start context attributes, end atomic activities, end context attributes and the threshold weight for the given activity were determined. For analyzing each of these complex activities the person was assumed to be in a sitting position before the start of the activity. For instance, for the complex activity of making food using microwave, as illustrated in Table 6, the activity analysis initiates with the process of the person moving towards the microwave and following the step by step sequence of performing the activity, based on the given context parameters and scenarios. The atomic activities and context attributes were assigned weights by probabilistic reasoning.

The respective atomic activities are: At1: Standing, At2: Walking Towards Microwave, At3: Loading Food In Microwave Bowl, At4: Setting The Time, At5: Turning on microwave, At6: Taking Out Bowl, At7: Sitting Back. The weights associated to these respective atomic activities are: At1: 0.10, At2: 0.12, At3: 0.14, At4: 0.15, At5: 0.25, At6: 0.18, At7: 0.06. The context attributes associated to these atomic activities are: Ct1: Lights on, Ct2: Kitchen Area, Ct3: Food Present, Ct4: Time settings working, Ct5: Microwave Present, Ct6: Bowl cool, Ct7: Sitting Area. The weights associated to these respective context attributes are: Ct1: 0.10, Ct2: 0.12, Ct3: 0.14, Ct4: 0.15, Ct5: 0.25, Ct6: 0.18, Ct7: 0.06.

As observed from the analysis shown in Table 6, the atomic activities At4, At5 and At6 have the highest weights so they are considered as the core atomic activities for this complex activity. The context parameters associated with these core atomic activities, Ct4, Ct5 and Ct6 are thus considered as the core context attributes. The atomic activities At1 and At2 and their associated context attributes Ct1 and Ct2 are concerned with the user getting up from the sitting position and initiating the process of this complex activity of making food using the microwave. Thus, they are identified as the start atomic activities and the start context attributes respectively. Similarly, the atomic activities At6 and At7 and their associated context attributes Ct6 and Ct7 relate to the user completing this activity and sitting down to enjoy the food. Thus, they are identified as the end atomic activities and end context attributes respectively.

**TABLE 1:** Analysis by CARALGO of the Complex Activity of Watching TV (WT).

| Complex Activity WCAtk (WT Atk) - WT (0.67) | |
|---|---|
| Weight of Atomic Activities WtAti | At1: Standing (0.15) At2: Walking towards TV (0.15) At3: Turning on the TV (0.25) At4: Fetching the remote control (0.15) At5: Sitting Down (0.08) At6: Tuning Proper Channel (0.12) At7: Adjusting Display and Audio (0.10) |
| Weight of Context Attributes WtCti | Ct1: Lights on (0.15) Ct2: Entertainment Area (0.15) Ct3: Presence of TV (0.25) Ct4: Remote Control Available (0.15) Ct5: Sitting Area (0.08) Ct6: Channel Present (0.12) Ct7: Settings working (0.10) |
| Core γAt and ρCt | At2, At3, At4 and Ct2, Ct3, Ct4 |
| Start AtS and CtS | At1, At2 and Ct1, Ct2 |
| End AtE and CtE | At5, At6, At7 and Ct5, Ct6, Ct7 |

International Journal of Recent Trends in Human Computer Interaction (IJHCI), Volume (9):Issue (1):2019   7

**TABLE 2:** Analysis by CARALGO of the Complex Activity of Using Laptop (UL).

| Complex Activity WCAtk (WT Atk) - UL (0.82) | |
|---|---|
| Weight of Atomic Activities WtAti | At1: Standing (0.10) At2: Walking Towards Laptop Area (0.15) At3: Turning on Laptop (0.28) At4: Typing log in password (0.23) At5: Sitting Down near Laptop (0.06) At6: Opening Required Application (0.10) At7: Connecting any peripheral devices like mouse, keyboard etc. (0.08) |
| Weight of Context Attributes WtCti | Ct1: Lights on (0.10) Ct2: Laptop Table (0.15) Ct3: Laptop Present (0.28) Ct4: Log-in feature working (0.23) Ct5: Sitting Area (0.06) Ct6: Required Application Present (0.10) Ct7: Peripheral devices (0.08) |
| Core γAt and ρCt | At2, At3 and Ct2, Ct3 |
| Start AtS and CtS | At1, At2, and Ct1, Ct2 |
| End AtE and CtE | At6, At7 and Ct6, Ct7 |

**TABLE 3:** Analysis by CARALGO of the Complex Activity of Listening to Subwoofer (LTS).

| Complex Activity WCAtk (WT Atk) - LTS (0.70) | |
|---|---|
| Weight of Atomic Activities WtAti | At1: Standing (0.13) At2: Walking Towards Subwoofer (0.18) At3: Turning on the Subwoofer (0.23) At4: Plugging the CD/DVD/Storage Device for Playing (0.16) At5: Adjusting Sound Controls (0.11) At6: Adjusting Display Controls (0.11) At7: Sitting Down (0.08) |
| Weight of Context Attributes WtCti | Ct1: Lights on (0.13) Ct2: Entertainment Area (0.18), Ct3: Subwoofer Present (0.23) Ct4: CD/DVD/Storage Device Present (0.16) Ct5: Sound Controls Working (0.11) Ct6: Display Controls Working (0.11) Ct7: Sitting Area (0.08) |
| Core γAt and ρCt | At2, At3, At4 and Ct2, Ct3, Ct4 |
| Start AtS and CtS | At1, At2 and Ct1, Ct2 |
| End AtE and CtE | At6, At7 and Ct6, Ct7 |

**TABLE 4:** Analysis by CARALGO of the Complex Activity of Using Washing Machine (UWM).

| Complex Activity WCAtk (WT Atk) - UWM (0.72) | |
|---|---|
| Weight of Atomic Activities WtAti | At1: Standing (0.08) At2: Walking Towards Machine (0.20) At3: Turning On Machine (0.25) At4: Pouring Detergent (0.08) At5: Loading Clothes (0.20) At6: Adjusting Timer (0.12) At7: Sitting Down (0.07) |
| Weight of Context Attributes WtCti | Ct1: Lights on (0.08) Ct2: Laundry Area (0.20) Ct3: Washing Machine present (0.25) Ct4: Detergent Available (0.08) Ct5: Presence of clothes (0.20) Ct6: Timer settings working (0.12) Ct7: Sitting Area (0.07) |
| Core γAt and ρCt | At3, At5 and Ct3, Ct5 |
| Start AtS and CtS | At1, At2 and Ct1, Ct2 |
| End AtE and CtE | At6, At7 and Ct6, Ct7 |

**TABLE 5:** Analysis by CARALGO of the Complex Activity of Cooking Food Using Kettle (CFWK).

| Complex Activity WCAtk (WT Atk) - CFWK (0.60) | |
|---|---|
| Weight of Atomic Activities WtAti | At1: Standing (0.10) At2: Walking Towards Kettle (0.16) At3: Loading Food Into Kettle (0.19) At4: Turning On Burner (0.12) At5: Adjusting Heat (0.09) At6: Adding spices in Food (0.11) At7: Stirring (0.07) At8: Turning Off burner (0.12) At9: Sitting Back (0.04) |
| Weight of Context Attributes WtCti | Ct1: Lights on (0.10) Ct2: Kettle Present (0.16) Ct3: Food to be cooked (0.19) Ct4: Burner Turning On (0.12) Ct5: Heat Settings (0.09) Ct6: Food spices (0.11) Ct7: Stirrer (0.07) Ct8: Burner Turning off (0.12) Ct9: Sitting Area (0.04) |
| Core γAt and ρCt | At2,At3 and Ct2,Ct3 |
| Start AtS and CtS | At1, At2 and Ct1, Ct2 |
| End AtE and CtE | At8, At9 and Ct8, Ct9 |

**TABLE 6:** Analysis by CARALGO of the Complex Activity of Making Food Using Microwave (MFUM).

| Complex Activity WCAtk (WT Atk) - MFUM (0.73) | |
|---|---|
| Weight of Atomic Activities WtAti | At1: Standing (0.10) At2: Walking Towards Microwave (0.12) At3: Loading Food In Microwave Bowl (0.14) At4: Setting The Time (0.15) At5: Turning on microwave (0.25) At6: Taking Out Bowl (0.18) At7: Sitting Back (0.06) |
| Weight of Context Attributes WtCti | Ct1: Lights on (0.10), Ct2: Kitchen Area (0.12), Ct3: Food Present (0.14), Ct4: Time settings working (0.15), Ct5: Microwave Present (0.25), Ct6: Bowl cool (0.18), Ct7: Sitting Area (0.06) |
| Core γAt and ρCt | At4, At5, At6 and Ct4, Ct5, Ct6 |
| Start AtS and CtS | At1, At2 and Ct1, Ct2 |
| End AtE and CtE | At6, At7 and Ct6, Ct7 |

**TABLE 7:** Analysis by CARALGO of the Complex Activity of Making Breakfast Using Toaster (MBUT).

| Complex Activity WCAtk (WT Atk) - MBUT (0.73) | |
|---|---|
| Weight of Atomic Activities WtAti | At1: Standing (0.10) At2: Walking Towards Toaster (0.12) At3: Putting bread into Toaster (0.15) At4: Setting The Time (0.15) At5: Turning off toaster (0.25) At6: Taking out bread (0.18) At7: Sitting Back (0.05) |
| Weight of Context Attributes WtCti | Ct1: Lights on (0.10), Ct2: Kitchen Area (0.12), Ct3: Bread Present (0.15), Ct4: Time settings working (0.15), Ct5: Toaster Present (0.25), Ct6: Bread cool (0.18), Ct7: Sitting Area (0.05) |
| Core γAt and ρCt | At3, At4, At5 and Ct3, Ct4, Ct5 |
| Start AtS and CtS | At1, At2 and Ct1, Ct2 |
| End AtE and CtE | At6, At7 and Ct6, Ct7 |

Nirmalya Thakur & Chia Y. Han

To study the patterns between these different complex activities with respect to time, the different time instants when each of them occurred were clustered using K nearest neighbor classification. During the process of clustering each cluster represented a different complex activity. These clusters were plotted to visualize and understand the patterns of these complex activities as well as analyze the sequence in which they were performed. This is shown in Figure 5.

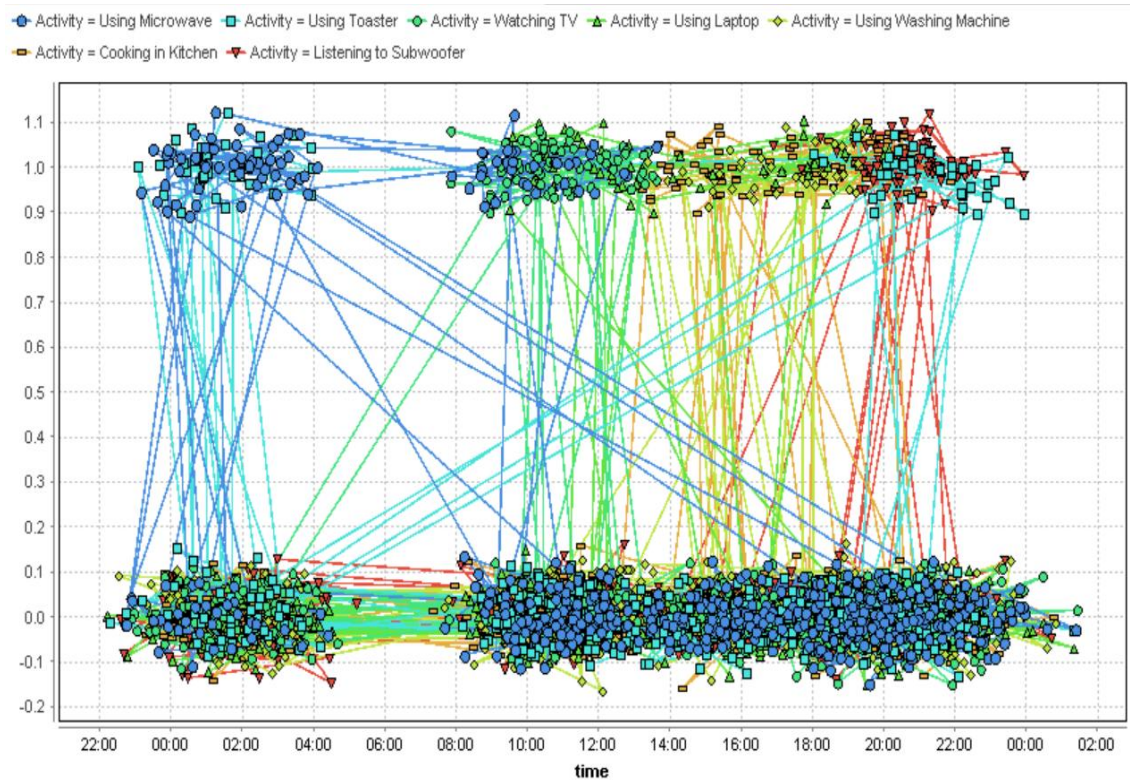

**FIGURE 5:** Cluster Based Analysis showing patterns and associations between different complex activities from the subset of the UK DALE dataset [27].

After analyzing through CARALGO, the different instances of these complex activities were analyzed through CABERA [25] to find the emotional response associated to each of their occurrences. Thereafter, a related work [26] was used to relate these emotional responses to the user experience associated to the different occurrences of these respective activities. This work [26] involves a random-forest based supervised learning model which uses the information about the emotional response of a complex activity to map it to a good or bad user experience.

This information was used to develop a complex activity recommendation system in RapidMiner [28] by considering the specific time instants when these activities were performed, the users affective state for each of these instances and the underlining user experience. This system is shown in Figure 6 and the flowchart of the same is shown in Figure 7. The input data to this system consisted of different complex activities retrieved from a subset of the UK DALE dataset [27]. This data was split into training set and test set as 70% training data and 30% test data. The test data was used to determine the performance accuracy of this system, which is discussed in the next section.

On running this system, for each activity performed by the user, it would evaluate the possibility of recommendation of all other activities by assigning confidence values to each of them. These confidence values indicated likeliness of the user performing each of those activities next, based on the time instant, affective state and user experience. A greater value of this confidence indicated a greater probability of that specific activity being performed next by the user. The

Nirmalya Thakur & Chia Y. Han

activity with the highest confidence value was recommended by the system for the given activity. The output of the system showed the current activity, the confidence values associated to all other activities and the recommended activity based on these confidence values. This is illustrated in Figure 8 and Figure 13, by screenshots of some of the output values, when the system modelled an "average user" and a "specific user" respectively.

In Figure 8, the output data consists of one attribute named "Activity" which is the actual activity that was performed by the user after the current activity. The next attribute named "prediction(Activity)" lists the activity recommended by the system. The next subsequent attributes list the confidence values indicating the degree of likeliness of the user performing the other activities for each given complex activity. These respective attributes are confidence(Using Microwave), confidence(Listening to Subwoofer), confidence(Watching TV), confidence(Using Laptop), confidence(Using Washing Machine), confidence(Cooking in Kitchen) and confidence(Using Toaster).

Similarly, in Figure 13, the output data consists of one attribute named "Activity" which is the actual activity that was performed by the user after the current activity. The next attribute named "prediction(Activity)" lists the activity recommended by the system. The next subsequent attributes list the confidence values indicating the degree of likeliness of the user performing the other activities for each given complex activity. These respective attributes are confidence(Sleeping), confidence(Watching TV in Spare Time), confidence(Showering), confidence(Eating Breakfast), confidence(Leaving), confidence(Eating Lunch) and confidence(Eating Snacks).

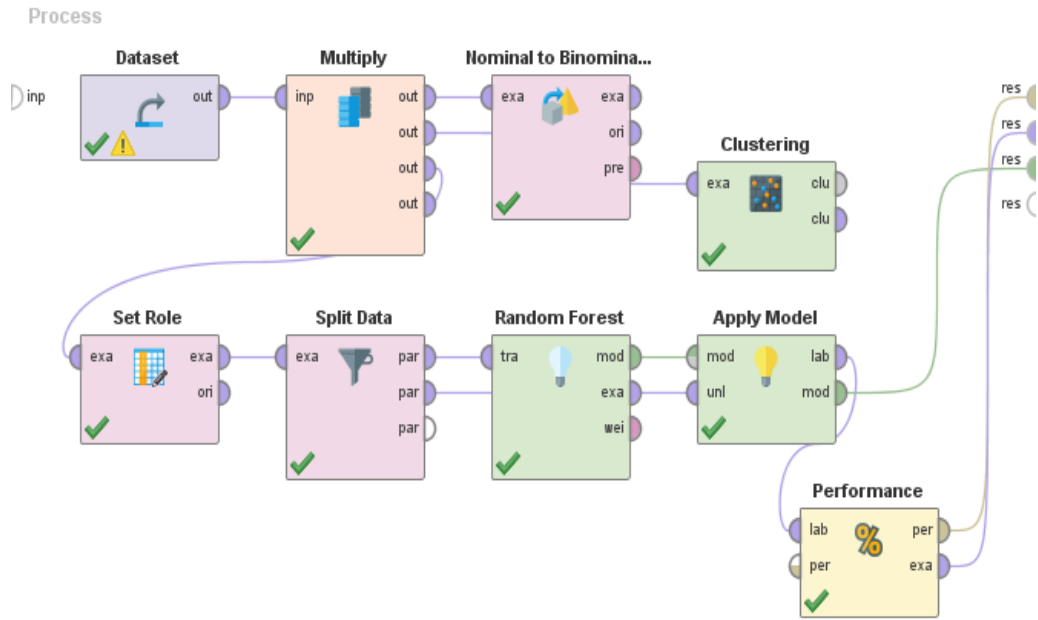

**FIGURE 6:** The system developed in RapidMiner for implementation of this framework.

Nirmalya Thakur & Chia Y. Han

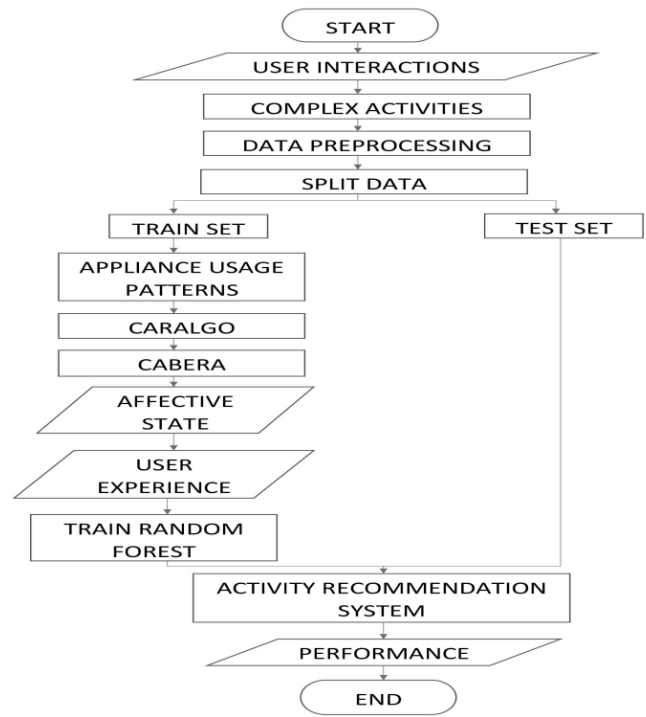

**FIGURE 7:** Flowchart of the system developed in RapidMiner for development of this Framework.

| Row No. | Activity | prediction(Activity) | confidence(... | confidence(... | confidence(... | confidence(... | confidence(... | confidence(... | confidence(... |
|---|---|---|---|---|---|---|---|---|---|
| 1 | Using Microwave | Using Microwave | 1 | 0 | 0 | 0 | 0 | 0 | 0 |
| 2 | Using Microwave | Using Microwave | 0.990 | 0.010 | 0 | 0 | 0 | 0 | 0 |
| 3 | Using Microwave | Using Microwave | 0.587 | 0 | 0.413 | 0 | 0 | 0 | 0 |
| 4 | Using Microwave | Using Microwave | 0.696 | 0 | 0.294 | 0.010 | 0 | 0 | 0 |
| 5 | Using Microwave | Using Microwave | 0.684 | 0 | 0.316 | 0 | 0 | 0 | 0 |
| 6 | Watching TV | Watching TV | 0.404 | 0 | 0.596 | 0 | 0 | 0 | 0 |
| 7 | Using Laptop | Using Laptop | 0 | 0 | 0.131 | 0.869 | 0 | 0 | 0 |
| 8 | Using Laptop | Using Laptop | 0.080 | 0 | 0.018 | 0.902 | 0 | 0 | 0 |
| 9 | Using Laptop | Using Laptop | 0 | 0 | 0 | 0.890 | 0 | 0.110 | 0 |
| 10 | Using Washing M... | Using Washing M... | 0 | 0 | 0 | 0.020 | 0.980 | 0 | 0 |
| 11 | Using Washing M... | Using Washing M... | 0 | 0 | 0 | 0 | 0.880 | 0.120 | 0 |
| 12 | Cooking in Kitchen | Cooking in Kitchen | 0 | 0 | 0 | 0 | 0.370 | 0.630 | 0 |
| 13 | Cooking in Kitchen | Cooking in Kitchen | 0 | 0 | 0 | 0 | 0.074 | 0.926 | 0 |
| 14 | Cooking in Kitchen | Cooking in Kitchen | 0 | 0 | 0 | 0 | 0.004 | 0.996 | 0 |

**FIGURE 8:** Screenshot of the output of the system which shows the recommended complex activities when modelling an "average user". The respective attributes are Activity, Prediction(Activity), confidence(Using Microwave), confidence(Listening to Subwoofer), confidence(Watching TV), confidence(Using Laptop), confidence(Using Washing Machine), confidence(Cooking in Kitchen) and confidence(Using Toaster).

**3.2 Implementation of the Framework to model a "specific user"**
To evaluate the performance of this framework on a "specific-user", the framework was then tested on a dataset which consisted of Activities of Daily Living (ADLs) performed by an elderly person (in the age range of 65-85) in the context of a smart home. This dataset was a result of

Nirmalya Thakur & Chia Y. Han

the work done by Ordóñez et al. [29]. This work consisted of developing a smart IoT-based environment which comprised of multiple sensors that were used to perform activity recognition to sense ADLs in a smart home, over 24 hours for a period of 22 days. It involved the use of ANN (Artificial Neural Network) and SVM (Support Vector Machines), within the framework of an HMM (Hidden Markov Model) to develop a learning model that analyzed different complex activity occurrences and recorded the same with time stamps. The work also compared the performance of this hybrid learning model with other learning models to uphold the efficacy of same in correctly understanding ADLs in a smart home setting.

The different ADLs that were a part of this dataset consisted of the complex activities of Sleeping, Showering, Eating Breakfast, Leaving for work, Eating Lunch, Eating Snacks and Watching TV in Spare Time. Multiple occurrences of these complex activities from this dataset with respect to time have been shown in Figure 9 for illustration. The same system (just the input was changed to this dataset) as shown is Figure 6, was used to implement this activity recommendation model and the flow of actions were also the same as shown in Figure 7. Similar to the methodology for modelling an "average user", this analysis involved studying the different occurrences of these complex activities with respect to time, the associated affective states of the user for each of these occurrences and the underlining user experience to build the complex activity recommendation system. The different instances of occurrences of the complex activities – Making Breakfast and Eating Lunch are shown in Figures 10-11. The CARALGO analysis of all the complex activities – Sleeping, Showering, Eating Breakfast, Leaving for work, Eating Lunch, Eating Snacks and Watching TV in Spare Time are shown in Tables 8-14.

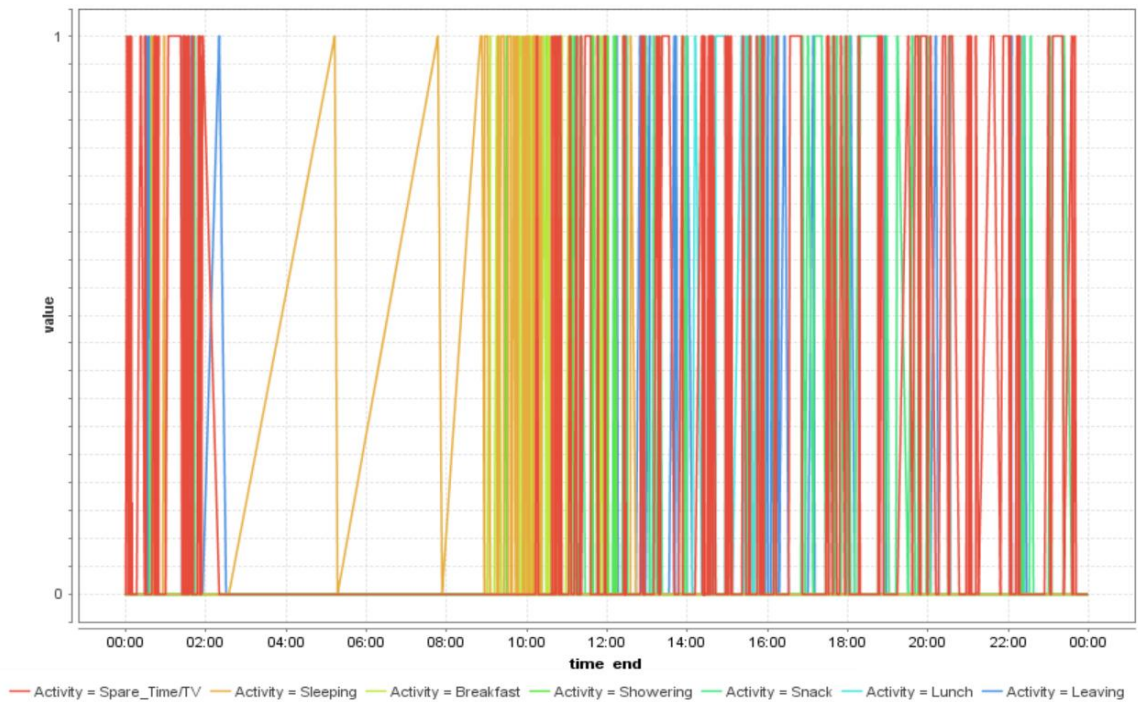

**FIGURE 9:** Different instances of activity occurrences from the dataset [29]. This includes the complex activities of Sleeping, Watching TV in Spare Time, Showering, Eating Breakfast, Leaving, Eating Lunch and Eating Snacks.

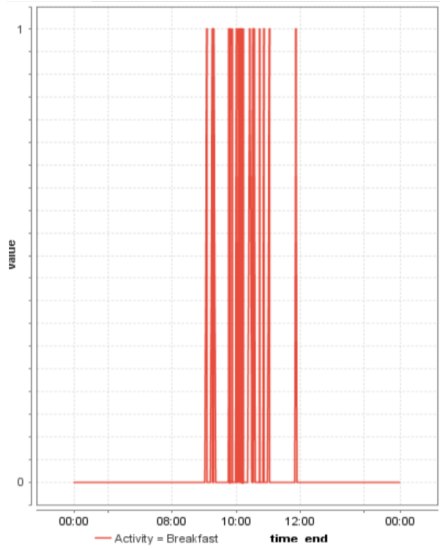 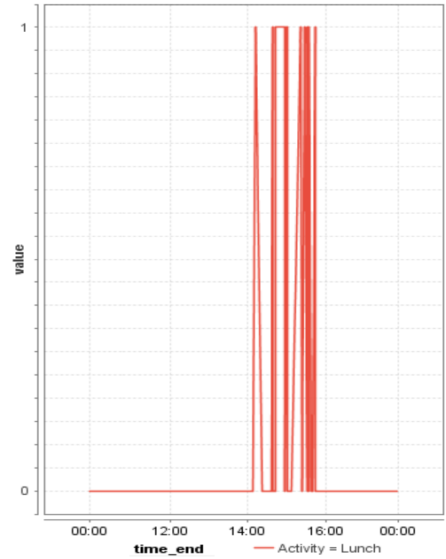

**FIGURE 10:** Multiple instances of the complex activity of Eating Breakfast.

**FIGURE 11:** Multiple instances of the complex activity of Eating Lunch.

**TABLE 8:** Analysis by CARALGO of the Complex Activity of Leaving (LV).

| | |
|---|---|
| Complex Activity WCAtk (WT Atk) - LV (0.68) | |
| Weight of Atomic Activities WtAti | At1: Standing (0.12) At2: Putting on dress to go out (0.32) At3: Carrying bag (0.30) At4: Walking towards door (0.13) At5: Going out of door (0.13) |
| Weight of Context Attributes WtCti | Ct1: Lights on (0.12), Ct2: Dress Present (0.32), Ct3: Bag present (0.30), Ct4: Exit Door (0.13), Ct5: Door working (0.13) |
| Core γAt and ρCt | At2, At3 and Ct2, Ct3 |
| Start AtS and CtS | At1, At2 and Ct1, Ct2 |
| End AtE and CtE | At4, At5 and Ct4, Ct5 |

**TABLE 9:** Analysis by CARALGO of the Complex Activity of Watching TV in Spare Time (WTV).

| | |
|---|---|
| Complex Activity WCAtk (WT Atk) – WTV (0.68) | |
| Weight of Atomic Activities WtAti | At1: Standing (0.10) At2: Walking Towards TV (0.30) At3: Turning on TV (0.28) At4: Tuning in the correct channel (0.15) At5: Starting to watch TV (0.17) |
| Weight of Context Attributes WtCti | Ct1: Lights on (0.10), Ct2: Entertainment Area (0.15), Ct3: татV present (0.15), Ct4: TV working (0.15), Ct5: Seating area (0.17) |
| Core γAt and ρCt | At2, At3 and Ct2, Ct3 |
| Start AtS and CtS | At1, At2 and Ct1, Ct2 |
| End AtE and CtE | At5 and Ct5 |

Nirmalya Thakur & Chia Y. Han

**TABLE 10:** Analysis by CARALGO of the Complex Activity of Making Breakfast (MB).

| Complex Activity WCAtk (WT Atk) - MB (0.73) | |
|---|---|
| Weight Of Atomic Activities WtAti | At1: Standing (0.10) At2: Walking Towards Kitchen (0.12) At3: Loading Food In Microwave (0.14) At4: Turning on Microwave (0.25) At5: Setting The Time (0.15) At6: Taking lout prepared breakfast (0.18) At7: Sitting down to eat (0.06) |
| Weight Of Context Attributes WtCti | Ct1: Lights on (0.10), Ct2: Kitchen Area (0.12), Ct3: Microwavable food Present (0.14), Ct4: Time settings (0.15), Ct5: Microwave Present (0.25), Ct6: Microwave Working (0.18), Ct7: Sitting down to eat (0.06) |
| Core γAt and ρCt | At4, At5, At6 and Ct4, Ct5, Ct6 |
| Start AtS and CtS | At1, At2 and Ct6, Ct7 |
| End AtE and CtE | At6, At7 and Ct6, Ct7 |

**TABLE 11:** Analysis by CARALGO of the Complex Activity of Eating Lunch (EL).

| Complex Activity WCAtk (WT Atk) – EL (0.72) | |
|---|---|
| Weight of Atomic Activities WtAti | At1: Standing (0.08) At2: Walking towards dining table (0.20) At3: Serving food on a plate (0.25) At4: Washing Hand/Using Hand Sanitizer (0.20) At5: Sitting down (0.08) At6: Starting to eat (0.19) |
| Weight of Context Attributes WtCti | Ct1: Lights on (0.08) Ct2: Dining Area (0.20) Ct3: Food present (0.25) Ct4: Plate present (0.20) Ct5: Sitting options available (0.08) Ct6: Food quality and taste (0.19) |
| Core γAt and ρCt | At2, At3, At4 and Ct2, Ct3, Ct4 |
| Start AtS and CtS | At1, At2 and Ct1, Ct2 |
| End AtE and CtE | At5, At6 and Ct5, Ct6 |

**TABLE 12:** Analysis by CARALGO of the Complex Activity of Eating Snacks (ES).

| Complex Activity WCAtk (WT Atk) – ES (0.72) | |
|---|---|
| Weight of Atomic Activities WtAti | At1: Standing (0.08) At2: Walking towards dining table (0.20) At3: Serving food on a plate (0.25) At4: Washing Hand/Using Hand Sanitizer (0.20) At5: Sitting down (0.08) At6: Starting to eat (0.19) |
| Weight of Context Attributes WtCti | Ct1: Lights on (0.08) Ct2: Dining Area (0.20) Ct3: Food present (0.25) Ct4: Plate present (0.20) Ct5: Sitting options available (0.08) Ct6: Food quality and taste (0.19) |
| Core γAt and ρCt | At2, At3, At4 and Ct2, Ct3, Ct4 |
| Start AtS and CtS | At1, At2 and Ct1, Ct2 |
| End AtE and CtE | At5, At6 and Ct5, Ct6 |

Nirmalya Thakur & Chia Y. Han

**TABLE 13:** Analysis by CARALGO of the Complex Activity of Going to Sleep (GTS).

| Complex Activity WCAtk (WT Atk) – GTS (0.72) | |
|---|---|
| Weight of Atomic Activities WtAti | At1: Standing (0.12) At2: Walking Towards Bed (0.23) At3: Turning Off lights (0.28) At4: Setting Alarm (0.22) At5: Using blanket (0.15) |
| Weight of Context Attributes WtCti | Ct1: Lights on (0.12) Ct2: Bed Present (0.24) Ct3: Light switch working (0.28) Ct4: Alarm working (0.22) Ct5: Blanket Present (0.15) |
| Core γAt and ρCt | At2, At3 and Ct2, Ct3 |
| Start AtS and CtS | At1, At2 and Ct1, Ct2 |
| End AtE and CtE | At4, At5 and Ct4, Ct5 |

**TABLE 14:** Analysis by CARALGO of the Complex Activity of Taking Shower (TS).

| Complex Activity WCAtk (WT Atk) – TS (0.72) | |
|---|---|
| Weight of Atomic Activities WtAti | At1: Standing (0.08) At2: Walking Towards Shower Room (0.20) At3: Carrying soap and/or shampoo (0.25) At4: Carrying Towel (0.20) At5: Turning on shower (0.14) At6: Turning off shower (0.13) |
| Weight of Context Attributes WtCti | Ct1: Lights on (0.08) Ct2: Shower room unoccupied (0.25) Ct3: Soap/Shampoo present (0.20) Ct4: Towel present (0.20) Ct5: Shower working (0.12) Ct6: Shower tap working (0.15) |
| Core γAt and ρCt | At2, At3, At4 and Ct2, Ct3, Ct4 |
| Start AtS and CtS | At1, At2 and Ct1, Ct2 |
| End AtE and CtE | At5, At6 and Ct5, Ct6 |

To study the patterns between these different complex activities with respect to time, the different time instants when each of them occurred were clustered using K nearest neighbor classification. During the process of clustering each cluster represented a different complex activity. These clusters were plotted to visualize and understand the patterns of these complex activities as well as analyze the sequence in which they were performed. This is shown in Figure 12.

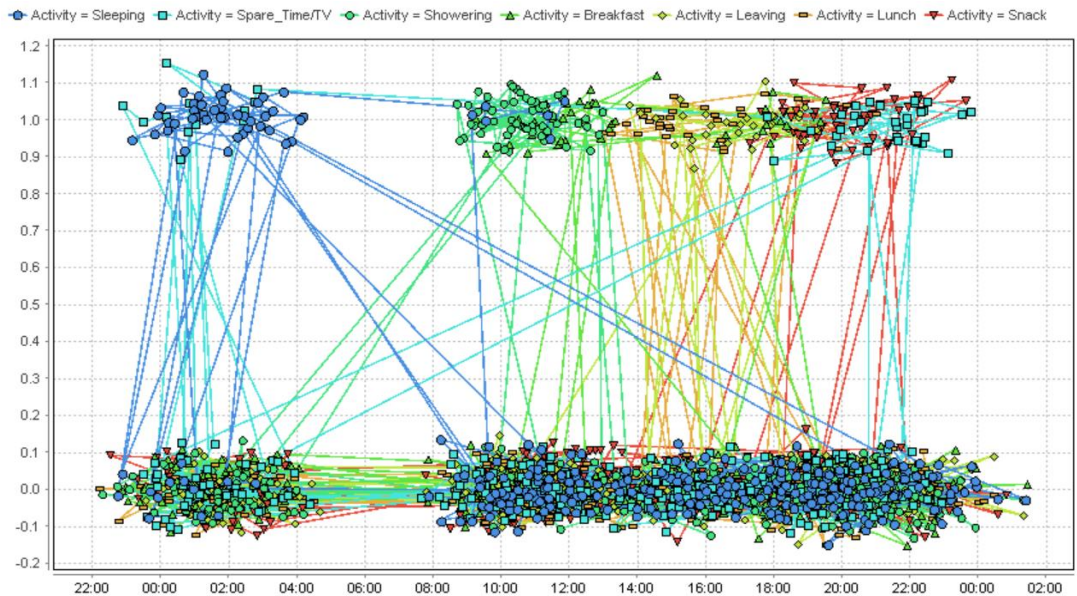

**FIGURE 12:** Cluster Analysis (KNN Nearest neighbor) of the different activities from the dataset [29] to analyze the patterns and associations amongst them.

Nirmalya Thakur & Chia Y. Han

| Row No. | Activity | prediction(A... | confidence(... | confidence(... | confidence(... | confidence(... | confidence(... | confidence(... | confidence(... |
|---|---|---|---|---|---|---|---|---|---|
| 1 | Spare_Time/... | Spare_Time/... | 0.097 | 0.903 | 0 | 0 | 0 | 0 | 0 |
| 2 | Sleeping | Sleeping | 0.903 | 0.097 | 0 | 0 | 0 | 0 | 0 |
| 3 | Sleeping | Sleeping | 0.983 | 0.017 | 0 | 0 | 0 | 0 | 0 |
| 4 | Showering | Showering | 0 | 0 | 1 | 0 | 0 | 0 | 0 |
| 5 | Showering | Showering | 0 | 0 | 1 | 0 | 0 | 0 | 0 |
| 6 | Showering | Showering | 0 | 0 | 1 | 0 | 0 | 0 | 0 |
| 7 | Breakfast | Breakfast | 0 | 0 | 0.074 | 0.926 | 0 | 0 | 0 |
| 8 | Breakfast | Breakfast | 0 | 0 | 0.323 | 0.677 | 0 | 0 | 0 |
| 9 | Breakfast | Leaving | 0 | 0 | 0 | 0.477 | 0.495 | 0.028 | 0 |
| 10 | Lunch | Lunch | 0 | 0 | 0 | 0.342 | 0.094 | 0.564 | 0 |
| 11 | Lunch | Lunch | 0 | 0 | 0 | 0.004 | 0.430 | 0.567 | 0 |
| 12 | Snack | Lunch | 0 | 0.020 | 0 | 0 | 0 | 0.794 | 0.186 |
| 13 | Snack | Snack | 0 | 0 | 0 | 0 | 0 | 0 | 1 |
| 14 | Snack | Snack | 0 | 0 | 0 | 0 | 0 | 0 | 1 |

**FIGURE 13:** Screenshot of the output of the system which showed the recommended activities when modelling a "specific user". The respective attributes are Activity, prediction(Activity), confidence(Sleeping), confidence(Watching TV in Spare Time), confidence(Showering), confidence(Eating Breakfast), confidence(Leaving), confidence(Eating Lunch) and confidence(Eating Snacks).

## 4. RESULTS AND DISCUSSION

The performance accuracies of this framework, in the form of confusion matrices, for modeling an "average user" and for modeling a "specific user" are shown in Figures 14 and 15 respectively. As observed from Figure 14, the overall performance accuracy of this model when modeling an "average user" is 62.59%. The respective sub-classes here being the different complex activities from the subset of the UK DALE dataset [27] that were used for developing this activity recommendation model. These different complex activities being Using Microwave, Using Toaster, Watching TV, Using Laptop, Using Washing Machine, Cooking in the Kitchen and Listening to the Subwoofer; with Watching TV having the highest sub-class precision of 77.78% and Using Toaster having the lowest sub-class precision of 47.62%.

accuracy: 62.59%

|  | true Using Mi... | true Using To... | true Watching... | true Using La... | true Using W... | true Cooking ... | true Listening... | class precisi... |
|---|---|---|---|---|---|---|---|---|
| pred. Using ... | 15 | 10 | 1 | 0 | 0 | 0 | 0 | 57.69% |
| pred. Using T... | 3 | 10 | 0 | 0 | 0 | 0 | 8 | 47.62% |
| pred. Watchin... | 6 | 0 | 21 | 0 | 0 | 0 | 0 | 77.78% |
| pred. Using L... | 0 | 0 | 0 | 7 | 4 | 2 | 0 | 53.85% |
| pred. Using ... | 0 | 0 | 0 | 4 | 8 | 1 | 0 | 61.54% |
| pred. Cookin... | 0 | 0 | 0 | 4 | 1 | 13 | 3 | 61.90% |
| pred. Listenin... | 0 | 1 | 0 | 0 | 0 | 4 | 13 | 72.22% |
| class recall | 62.50% | 47.62% | 95.45% | 46.67% | 61.54% | 65.00% | 54.17% |  |

**FIGURE 14:** Confusion Matrix showing the performance accuracy of the system when it modeled an "average user" from the subset of the UK DALE dataset [27].

accuracy: 73.12%

| | true Sleeping | true Spare_Ti... | true Showering | true Breakfast | true Leaving | true Lunch | true Snack | class precisi... |
|---|---|---|---|---|---|---|---|---|
| pred. Sleeping | 12 | 2 | 1 | 0 | 0 | 0 | 0 | 80.00% |
| pred. Spare_... | 0 | 6 | 0 | 0 | 0 | 0 | 3 | 66.67% |
| pred. Shower... | 3 | 0 | 16 | 0 | 0 | 0 | 0 | 84.21% |
| pred. Breakfast | 0 | 0 | 0 | 8 | 0 | 0 | 0 | 100.00% |
| pred. Leaving | 0 | 0 | 0 | 3 | 6 | 1 | 0 | 60.00% |
| pred. Lunch | 0 | 0 | 0 | 0 | 3 | 7 | 1 | 63.64% |
| pred. Snack | 0 | 6 | 0 | 0 | 0 | 2 | 13 | 61.90% |
| class recall | 80.00% | 42.86% | 94.12% | 72.73% | 66.67% | 70.00% | 76.47% | |

**FIGURE 15:** Confusion Matrix showing the performance accuracy of the system when it modeled a "specific user" - an elderly person (aged in range of 65-85 years) from the dataset [29].

To evaluate the efficacy of this framework for modelling a "specific user", the system was tested on a dataset [29] which consisted of Activities of Daily Living (ADLs) performed by a "specific user" (elderly person, age group: 65-85 years) in the context of a smart home. As can be observed from Figure 15, the performance of the model significantly increased in this scenario. The respective sub-classes were again the different complex activities from the dataset [29] that was used for developing this recommendation model. These different complex activities being Sleeping, Showering, Eating Breakfast, Leaving for work, Eating Lunch, Eating Snacks and Watching TV in Spare Time. In this case, the complex activity of Eating Breakfast had the highest sub-class precision, with its value being as high as 100% and the complex activity of Leaving for Work had the lowest sub class precision with a value of 60%.

For modeling an "average user", the model took into consideration multiple traits, different interaction patterns, varying affective states and different user experiences of all the users who performed those complex activities at different time instants. The recommendation of complex activities based on this "average user" modelling significantly varied in comparison to the interaction patterns of the "actual user" and this explains the reason for the low performance accuracy for that specific scenario. On the other hand, when the system was implemented for a "specific user" the demographics, interaction patterns and context parameters surrounding the user were known. This allowed the system to train itself based on the interaction styles, affective states and user experience that were specific to the given user. Thus, the complex activity recommendations made by the system mostly matched the interaction patterns of the user under consideration, which attributed to a greater performance accuracy of the system.

The application of activity centric computing in the context of smart homes has been of significant focus to researchers in the field of human-computer interaction. However, the approaches for activity recognition proposed by recent researchers [11-16], as discussed in this paper, were applied in specific settings and were tested on specific groups of users accessible to the respective authors. Research works in this field have shown that the diversity in users tend to have an impact on the way people interface with technology and behave in any given setting. So, to make activity recognition in the context of smart homes more reliable and suited to varying diversities in users, it is essential that activity recognition approaches do not get affected by universal diversity. The activity recognition approach, CARALGO [24] used in this work, is based on probabilistic reasoning and analyzes the probability of occurrence of different atomic activities with respect to their context attributes to infer about the occurrence of a complex activity. CARALGO [24] is not confined to a specific setting and can easily be implemented in any IoT-based environment. Also, CARALGO [24] has also been found to have a very high-performance accuracy of 88.5% in correctly analyzing human activities; which is much higher than the performance accuracies of the above-mentioned approaches [11-16] for activity analysis.

Nirmalya Thakur & Chia Y. Han

Speaking in terms of elderly people, who often have face Mild Cognitive Impairment (MCI), leading to forgetfulness, memory problems and cognitive issues, it is essential for the future of smart home intelligent technologies to not only analyze their behavior but also to recommend activities for development of an assistive environment to support independent living. Therefore, in addition to activity recognition, this framework presents an approach for activity recommendation in a smart home. The previous works on activity recommendation were mostly based on analyzing the user's daily routine to recommend tasks and actions. Research in this field has shown that a user may not always follow his or her daily routine. This can be due to various reasons out of which emotion, belief, desire and intention [30] of the user at that specific instant play a significant role in deciding the nature of activities that the user intends to perform next. Thus, it is essential for adaptive and assistive systems to be able to analyze the users affective state in addition to the users daily routine, before recommending activities to the user. This proposed framework uses CABERA [25] – to analyze the users affective state and uses clustering methods to develop relationships between the different activities performed by the user on a daily basis, so that while recommending a complex activity to the user both these aspects, i.e. the users affective state as well as the users daily routine are taken into consideration.

In addition to the above, the recent works in this field of activity recommendation in a smart home, as discussed earlier [17-23] which have focused on activity recommendation in specific settings, have modelled an "average user" by analyzing user interaction patterns of multiple users. The essence of developing an assistive environment for elderly people to help them perform their daily routine tasks is to create adaptive technologies that can foster their independent living in the context of their ADLs. To achieve the same, it is important for intelligent systems to be able to adapt according to the varying user interaction patterns leading from the diversity in elderly people. Therefore, this work proposes the framework for such a personalized intelligent assistant that can adapt with respect to the diversities in elderly people and recommend complex activities to help them have a better quality of life. The results presented in this work compare the performance of an activity recommendation system modelling an "average user" and an activity recommendation system modelling a "specific user". It is observed that there is a 16.8% increase in the performance accuracy of the system when it models a "specific user" and this upholds the relevance of the proposed framework and its implementation in the future of smart homes and smart environments for helping elderly people perform ADLs. The performance accuracy of 73.12%, achieved by this framework for recommending complex activities to a "specific user" is also higher than majority of the activity recommender systems as discussed in [17-23].

## 5. CONCLUSION AND FUTURE WORK

To increase the assistive nature of Affect Aware Systems in the context of smart and connected IoT-based living spaces for improving the quality of life experienced by elderly people, it is not only essential to analyze user behavior, but it is also important to help them augment their performances in the context of their day to day goals. Cognitive issues, weakened memory, disorganized behavior and even physical limitations [2, 3] are some of the major problems that elderly people face with increasing age.

Recent researches [11-20] in this field which have focused on providing technology-based solutions to address these needs of elderly people have these limitations – (1) Majority of these systems [11-16] have focused on multimodal ways of activity recognition with limited scope of augmenting the performance of the user in the given context; (2) The few task recommendation systems that have been developed [17-23] model an "average user". The traits and characteristics of this "average user" could be significantly different from a "specific user" owing to the user diversity; (3) The models are mostly applicable in specific environments, for instance a hospital room [17] and have limited scope for their implementation in any given context.

It is essential to address this "gap" in making technology-based solutions more relevant by improving their ability to address the diversity in users and adapt according to the specific needs of elderly people, to enhance their quality of life and augment their performances in the context of their day to day goals. This paper, therefore, proposes a framework for development of a Personalized Intelligent Assistant that can analyze the tasks performed by elderly people in a smart home environment and recommend activities based on their daily routine, affective states and the underlining user experience, by adapting to the specific interaction styles of the "specific user" being modelled.

The proposed framework has been tested on a couple of datasets to uphold the relevance of the same. The presented results discuss the performance of this framework for modeling an "average user" and a "specific user". The comparison of the performance characteristics of the two systems, upholds the efficacy of this model to address and adapt to the needs of a "specific user" in a given smart environment. To the best knowledge of the authors, no prior work has been done in this field that integrates the concept of affect aware systems with activity centric computing to develop a framework that can recommend complex activities to elderly people in the context of a smart home environment to help them perform ADLs.

Future work along these lines would involve deployment of multiple sensors to set up a smart and connected IoT-based environment. Thereafter this Personalized Intelligent Assistant would be implemented in real time to analyze the effectiveness of the same in recommending ADLs to elderly people to improve their quality of life and enhance user experiences, by modelling each user as a "specific user" in the given context.